\documentstyle[sprocl]{article}

\input{psfig}

\def\be{\begin{equation}}
\def\ee{\end{equation}}
\def\bea{\begin{eqnarray}}
\def\eea{\end{eqnarray}}

\begin{document}

\begin{flushright}
LU TP 00-26\\
hep-lat/0006006\\
\today
\end{flushright}
\vskip0.5cm

\title{DO INSTANTONS OF THE CP(N-1) MODEL MELT?}

\author{M.~Maul}

\address{Department of Theoretical Physics,
Lund University
S\"olvegatan 14A,\\
S - 223 62 Lund, Sweden\\
E-mail: maul@thep.lu.se}

\author{D.~Diakonov}
\address{Nordic Institute for Theoretical Physics (NORDITA), \\
Blegdamsvej 17, DK-2100 Copenhagen, Denmark\\
E-mail: diakonov@nordita.dk}


\maketitle
\abstracts{
In the two-dimensional $CP^{N-1}$ model one can parametrize exact
many-instanton solutions via $N$ `constituents' (called `zindons').
This parameterization allows, in principle, a complete `melting'
of individual instantons. The model is therefore well suited
to study whether dynamics prefers a dilute or a strongly
overlapping ensemble of instantons. We study the statistical mechanics
of instantons both analytically and numerically. We find that at $N=2$
the instanton system collapses into zero-size instantons. At $N=3,4$
we find that well-isolated instantons are dynamically preferred
though 15-25\% of instantons have a considerable overlap with others.
}

\section{Introduction}
Instantons, the specific fluctuations of the gluon field, carrying
topological charge, play an important role in explaining many
features  of QCD, like the spontaneous breaking of  chiral symmetry
\cite{Diakonov:1986eg,Schafer:1998wv}. Furthermore, the
instanton vacuum calculations are capable of providing the
non-perturbative input to a variety of observables
in Deep Inelastic Scattering (DIS)
like polarized and unpolarized
parton distributions \cite{Diakonov:1996sr},
two-hadron distribution amplitudes
skewed parton distributions and \cite{Polyakov:1999td},
higher-twist matrix elements \cite{Dressler:1999zi}
and other observables.

The role of instantons in the confinement phenomenon
is still not clear. In general, a rigorous proof of a linear confining
potential between static probe quarks in a 4-dimensional pure
Yang--Mills theory  from  first principles is still missing
\cite{Haymaker:1999cw}, while the extraction of that potential from
the current lattice data is subject to large systematic uncertainties
\cite{Diakonov:2000rk}.

It has been noticed some time ago \cite{proc95} that
an infinitely rising linear potential may be achieved if the
instanton size distribution falls off as $\nu(\rho)\sim 1/\rho^3$
at large $\rho$. Such a regime would mean that large instantons overlap,
and that the widely used sum ansatz of single instanton solutions
is not too meaningful.
If instantons are of any relevance for confinement, it cannot be seen
in the dilute-gas approximation.
Unfortunately, the true multi-instanton solution is not available
in QCD: the long-known ADHM multi-instanton solution \cite{Atiyah:1978ri}
is not an explicit one.

This motivates to investigate the overlap of instantons in a theory
which is more simple than QCD. Such a theory is the
two-dimensional CP$^{N-1}$ model. The model
contains asymptotic freedom, confinement and instantons
whose explicit form is known for any topological number and any number of
colors $N$ \cite{GP:1978,DAdda:1978uc}. The model is solvable
at large $N$ \cite{DAdda:1979kp,Witten:1979bc}
and, most important, the true multi-instanton measure of the
theory is known analytically \cite{Fateev:1979xn,Berg:1979uq}.

This paper reports on some of the results of our study of the
statistical mechanics of instantons in the $CP^{N-1}$ model,
by combining analytical and numerical methods
\cite{Diakonov:2000ae}.
Our conclusion is that, though the bulk of instantons appears
to be well isolated, some 15-25\% of them have a significant
overlap.
\section{The CP$^{N-1}$ model}
The CP$^{N-1}$ model is defined in two dimensions which can be
represented by the complex plain. The dynamical variables are
the $N$ complex fields $u_A,\; A=1,\dots,N$, which are normalized
to unity:
\begin{equation}
u_A = \frac{v_A}{|v|};\quad |v|^2 = \sum_{A=1}^{N} |v_A|^2\;.
\end{equation}
We shall call the index $A$ `color' in analogy to QCD.
From the fields $u_A$ a vector potential $A_\mu$ can be constructed:
\begin{equation}
A_\mu = \frac{i}{2}
\left(u_A \partial_\mu u_A^* - u_A^* \partial_\mu u_A \right),
\quad (\mu =1,2)
\;.
\end{equation}
The theory is defined by the partition function:
\begin{equation}
{\cal Z} = \int\!
{\cal D}u_A(x)
{\cal D}u^*_A(x)
{\cal D}A_\mu(x)\delta(|u|^2-1)
\exp\left(-\frac{1}{g^2}
\int\!d^2x|\nabla_\mu u_A|^2\right),
\end{equation}
with the covariant derivative being

\begin{equation}
\nabla_\mu = \partial_\mu -i A_\mu\;.
\end{equation}
The fact that the fields $u_A$ are normalized to unity makes
the theory non-linear. The theory possesses the Abelian gauge
invariance. From the vector potential $A_\mu$
a topological charge density $q_T(x)$
can be defined as:
\begin{equation}
q_T(x) = \frac{1}{4\pi} \epsilon_{\mu\nu} F_{\mu\nu}\,, \quad
F_{\mu\nu} = \partial_\mu A_\nu - \partial_\nu A_\mu\;.
\end{equation}
Here $\epsilon_{\mu\nu}$ is the antisymmetric tensor, i.e. $\epsilon_{12} =1,
\epsilon_{21} =-1$, and 0 for the other two index combinations.
The multi--instanton (multi--anti-instanton) solution of the theory
is known exactly \cite{GP:1978,DAdda:1978uc} and can be expressed
in terms of the unnormalized fields $v_A$ up to an inessential
constant as a product of simple monomials:
\begin{eqnarray}
{\rm \bf instantons}: \quad v_A
&=&\prod_{i=1}^{N_+} (z-a_{Ai}); \quad z = x+iy \;,
\nonumber \\
{\rm \bf anti-instantons}: \quad v_A
&=&\prod_{j=1}^{N_-} (z^*-b^*_{Aj}); \quad z^* = x-iy \;.
\end{eqnarray}
$N_+$ is the number of instantons and $N_-$ the number
of anti-instantons.
A single instanton solution is therefore given by a single
monomial and characterized by $N$ 2-dimensional points $a_A$,
which are called `instanton zindons'. In the same way the
2-dimensional coordinates $b_A$ are called the positions of
`anti-instanton zindons'. The word
zindon is Persian or Tadjik and means `prison' or `castle'.
There are, thus, $N$ types of `colors' of instanton zindons
(denoted by $a_A$) and $N$ types of anti-instanton zindons
(denoted by $b_A$). It is essential, that
the true multi-instanton solution is a product and not a sum of
single-instanton solutions. As in QCD, single instanton solutions
show up as well defined peaks in the topological charge density:
\begin{eqnarray}
v_A &=& (z-a_A) \;\longrightarrow \; q_T(x) = \frac{1}{\pi}
\frac{\rho^2}{((x-x_0)^2 +\rho^2)^2}\; ,
\label{qt}
\end{eqnarray}
where $x_0$ is the instanton center coinciding with the center
of mass of $N$ zindons of different `colors' and $\rho$, the instanton
size, is given by the spatial dispersion of zindons:
\begin{eqnarray}
x_0 &=& \frac{1}{N} \sum_A a_A; \quad \
\rho^2 = \sum_A \frac{1}{N}| x_0-a_A|^2 \;.
\label{single}
\end{eqnarray}
The corresponding single anti-instanton topological charge density
has the same form, but with a negative sign, so it forms a local minimum
in the topological charge density. For the combination of multi-instantons
and multi-anti-instantons one conventionally uses the product
ansatz \cite{Bukhvostov:1981sn}:
\begin{equation}
v_A = \prod_{i=1}^{N_+} (z-a_{iA})\prod_{j=1}^{N_-} (z^*-b^*_{jA}) \;.
\label{ansatz}\end{equation}
Naturally, it is not an exact solution (it becomes such only in the
limit of large separations between instanton and anti-instanton zindons),
therefore the action computed on this ansatz is not a sum of the individual
actions. The corresponding interaction of instantons and anti-instantons
formulated in terms of zindons has been found in Ref.~\cite{Diakonov:2000ae},
see the factor $w_{ab}$ below.
Combining it with the known multi-instanton ($w_a$) and
multi-anti-instanton ($w_b$) weights \cite{Fateev:1979xn,Berg:1979uq}
describing the interaction of `same-kind' zindons, one
writes the partition function in the form of statistical mechanics
of interacting particles ($N$ kinds of instanton zindons
and $N$ kinds of anti-instanton zindons):
\begin{equation}
Z = \sum_{N_++N_-}
\frac{e^{i\theta N_+}}{(N_+!)^N}
\frac{e^{-i\theta N_-}}{(N_-!)^N}
\int {\cal D} a {\cal D} b
\Lambda^{2N(N_++N_-)} w_a w_b w_{ab}\;.
\label{partf}\end{equation}
%
%
$\Lambda$ is the only dimensional constant of the theory and
can be set to 1. In the parameterization
of ref. \cite{Diakonov:2000ae} $w_a$ is given by:
\begin{eqnarray}
w_a &=& \exp \left[\sum_{i<i'}^{N_+}\sum_A \ln((a_{Ai}-a_{Ai'})^2\Lambda^2)
                -\frac{N}{2} \sum_{i,i'}^N
                 \ln \left[\sum_{A<B}^N(a_{Ai}-a_{Bi'})^2\Lambda^2 \right]
                 \right]
\nonumber \\
&& \times \exp\left[ N \frac{N_+(N_+-1)}{2}\ln \frac{N(N-1)}{2}\right]\;.
\end{eqnarray}
The corresponding weight for the anti-zindon interaction
$w_b$ is defined similarly. The instanton--anti-instanton interaction
is described by the factor \cite{Diakonov:2000ae}:
\begin{eqnarray}
w_{ab} &=& \exp\left\{ 2\beta \sum_{i=1}^{N_+}
\sum_{i'=1}^{N_-} \sum_{A,B}^N {\cal P}_{AB}
\ln \left[ (a_{Ai}-b_{Bj})^2 \Lambda^2 \right]\right\},
\nonumber \\
&&  {\cal P}_{AB} = \left\{ \begin{array}{cc}
                             \frac{N-1}{N}; & A=B \\ & \\
                             -\frac{1}{N}  ; & A\neq B
                             \end{array} \right. \;.
\end{eqnarray}
$\beta = 2\pi/(g^2 N)$ is the coupling between instantons and
anti-instantons. The partition function
describes two systems of zindons, namely
instanton and anti-instanton ones, experiencing logarithmic
interactions, whose strength is $N-1$ times stronger
for same-color zindons than for  different-color zindons.
One has attraction for zindons/anti-zindons of different
color and repulsion for zindons/anti-zindons of the same color.
At $N=2$ corresponding to the $CP^1=O(3)$ model one can think of
the ensemble as of that of $e^+,e^-,\mu^+,\mu^-$ particles
\cite{Bukhvostov:1981sn}. The interaction of opposite-kind
zindons are suppressed by an additional factor $\beta = 2\pi/(g^2N_c)$.
Since it is a classical system and not a quantum-mechanical one
where stable atoms do exist, such an ensemble,
i.e. the CP$^1$ model is unstable, as we will see in the next section.
\section{Instanton size distribution}
The multi-instanton ansatz (\ref{ansatz}) allows for a complete
`melting' of instantons. Indeed, if zindons $a_A$ and $b_A$ are evenly
distributed in space individual instantons loose any meaning.
In principle, another scenario could take place: a clustering
of N-plets of zindons of $N$ different `colors' into
`color-neutral' objects. If such clusters are well isolated from
other color-neutral clusters, they form well-separated
or dilute instantons. We recall that a single instanton consists
of $N$ zindons $a_A$ with different colors $A$. Which scenario takes
place in reality is a matter of the dynamics of the ensemble given
by the partition function (\ref{partf}).

The partition function (\ref{partf}) describing the
instanton--anti-instanton ensemble in the zindon parameterization
has been simulated with a Metropolis algorithm in
Ref.~\cite{Diakonov:2000ae}. One of the main objectives
has been to find the size distribution of instantons.

The basic question is how to identify instantons and anti-instantons.
This is also a serious problem for lattice QCD (see
e.g.~\cite{Negele:1999ev}).
In the case of the CP$^{N-1}$ model we can
compare two ways of extracting the instanton content. The first one,
which we call `geometrical', is inspired by the zindon parameterization.
Given a configuration of zindons $a_{Ai}$ on the plain being
at the thermodynamical equilibrium according
to the partition function $Z$ one can group them into instantons
using the following procedure:
\begin{itemize}
\item Take the group of $N$ zindons of $N$ different colors,
which has the smallest dispersion
      $\rho^2 = \frac{1}{N}\sum_A |a_A -x_0|^2$
      out of the  ensemble
      and call this group  an instanton of size $\rho$
located at the position $x_0 = \frac{1}{N} \sum_A a_A$.
\item From the rest of the ensemble take out the next
      group of $N$ zindons with the smallest dispersion $\rho$,
      and so on until the whole ensemble
      has been grouped into instantons (and anti-instantons).
\end{itemize}
This `geometrical' identification
of instantons assumes that the overlap does not affect much the peak
structure of the topological charge density.

The second way of looking at the instanton size distribution is through
the topological charge density $q_T(x)$. To that end we compute
$q_T$ on a grid from an equilibrium distribution of zindons
obtained from running the Metropolis algorithm.
The grid size limits the resolution
of small size fluctuations, but not of large size ones.
The method used to identify instantons is the following:
\begin{itemize}
\item Find a local maximum $x_0$, i.e. a grid point where
the topological charge is larger than the one of the eight surrounding
neighbors on the 9-plet centered at this grid point. In the spirit
of the single instanton solution take then the first approximation for
the instanton size to be $\rho_0^2  = 1/(\pi q_T(x_0))$.
\item Interpolate the topological charge density on the 9 plet quadratically
and calculate the two curvatures $\lambda_1$ and $\lambda_2$. If
they are both negative then the local maximum is confirmed
and we obtain at the same time
two further estimates for the density via
$\rho_i^2 = \sqrt{4/(\pi|\lambda_i|)},\;i=1,2$.
\item
If all three values $\rho_0,\rho_1,\rho_2$ are smaller than $L/2$
where $L$ is the size of the box where we place the ensemble,
then the local maximum is accepted to be an instanton and the
size is given by the geometric mean of all three estimates, i.e. by
$\rho = (\rho_0\rho_1\rho_2)^{1/3}$.
\end{itemize}

\begin{figure}[t]
\centerline{\psfig{figure=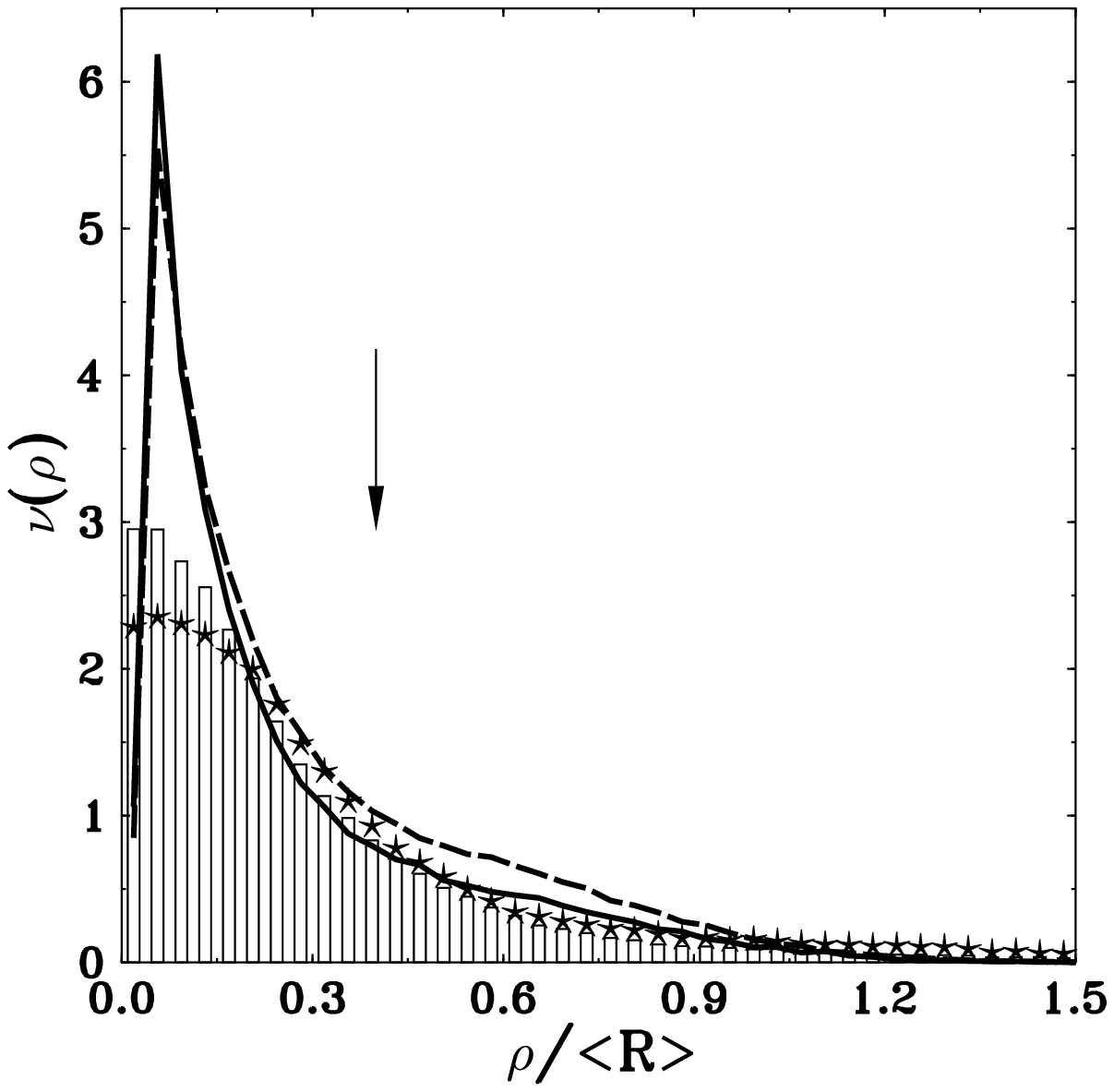,width=6cm}
            \psfig{figure=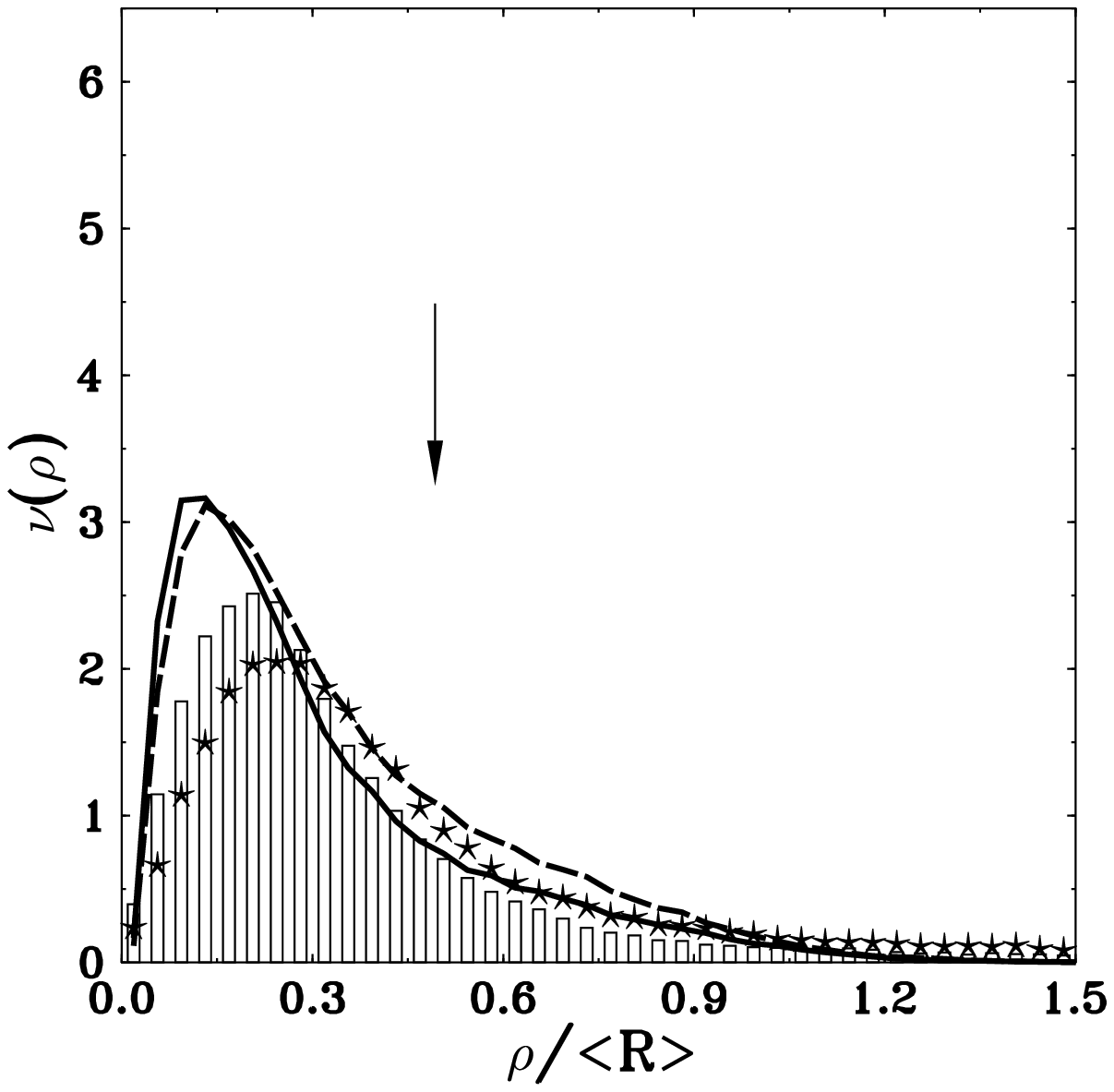,width=6cm}}
\caption{Instanton size distributions
are displayed for $N=3$ (left), and $N=4$ (right)
for $N_+=N_-=8$.
The histograms show the `geometric' size distributions, and
the solid lines the size distributions seen by the
`lattice' method, using a $100\times100$ grid.
The instanton--anti-instanton coupling constant
for the solid lines and the histograms is
$\beta=0$. The stars and the dashed lines show
the `geometric' and `lattice' size distributions for the case of
$\beta =0.5$.
Instanton sizes are plotted in units of the average separation
$\langle R\rangle$. The arrows show the maxima of the `geometric' 
size distributions obtained in the case of purely random space
distribution of zindons.}
\label{sizedist}
\end{figure}

Fig.~\ref{sizedist}, which has been taken from
\cite{Diakonov:2000ae}, shows the size distribution of instantons for
$N=3$ and $N=4$. In the case $N=2$ the zindons tend to condense
into color neutral pairs and the ensemble collapses, so from that
point of view the theory does
not exist at all for $N=2$. If one disregards the interaction of
instantons with anti-instantons then the attraction of different-color
zindons leads to a decrease of the average size of the instantons.
This can be seen by comparing the maximum of the size distribution for
$\beta=0$ with the maximum for an interaction-less purely random
distribution of zindons, which is represented by the arrows.

Switching in instanton--anti-instanton interactions, i.e.
moving from
$\beta=0$ [histogram/solid line] to
$\beta=0.5$ [stars/dashed line]
one observes that instantons are on the average shifted to
larger sizes, however  the effect is rather weak.
The effect of the size shrinkage owing to the multi-(anti)instanton
weight $w_{a(b)}$ is much more pronounced.

Remarkably, one observes a large discrepancy
between the instantons identified by the `lattice' method versus
the `geometric' method at small instanton sizes. The discrepancy
is prominent at $N=3$ but becomes considerably smaller
at $N=4$, so that a tendency is visible that it may die out
as $N$ is increased. This phenomenon of unphysical small size
fluctuations is similar to the well known `dislocation' phenomenon
observed in lattice studies \cite{Berg:1981er,Luscher:1982tq}.

For instantons with a size larger than half of the average separation,
i.e., $\rho/\langle R \rangle > 0.5$ one can say that the overlap
becomes essential. This is the case for $15$ -- $25\%$ of instantons,
displayed in the figure.
So one can say that the dilute gas ansatz
is justified for many purposes, but that there is a considerable amount
of instantons where the overlap cannot be neglected.

\section{Summary and conclusions}

We have formulated the statistical mechanics
of instantons and anti-instantons in the $d=2$ $CP^{N-1}$ model
in terms of their `constituents' which we call `zindons'.
We have derived the interactions of same-kind and opposite-kind
zindons for arbitrary $N$.

Though the zindon parameterization of instantons and of their interactions
allow for complete `melting' of instantons and is quite opposite in spirit
to dilute gas Ans\"atze, we observe that zindons, nevertheless, tend to
form `color-neutral' clusters which can be identified with well-isolated
instantons. This effect is due to a combination of two different
factors both supporting clustering. One factor is the interactions:
same-color zindons are strongly repulsive while different-color zindons
are attractive. The second factor is pure geometry: even with a purely
random distribution of zindons in space the probability to combine
$N$ zindons into a neutral cluster smaller than the average separation
is quite sizeable. Both these factors are expected to be even stronger
in four dimensions appropriate for the Yang-Mills instantons.

Despite an apparent tendency for clustering of zindons into
well-isolated instantons, there always exist a portion of
instantons which are strongly overlapping with the others.
Depending on what one calls a `strong' overlap we estimate the
portion of such instantons to be about 15-25\%. If a similar
effect takes place in the Yang--Mills case, it means that
there are long-range color correlations, which might be relevant
to confinement. At the same time if the bulk of instantons are well
separated (as we have found for the $CP^{(N-1)}$ model) it would
explain the success of instantons in describing physics related to
chiral symmetry breaking.

\section*{Acknowledgments}
One of us (M.M.) thanks the Knut and Alice Wallenberg Foundation for
financial support.

\end{document}